\documentclass[a4paper,12pt]{iopart}

\usepackage{graphicx,color,amssymb}

\newcommand*{\bto}{Bi$_2$Ti$_2$O$_6$O$^\prime$}

\newcommand*{\pro}{Pb$_2$Ru$_2$O$_6$O$^\prime_{0.5}$}
\newcommand*{\prosix}{Pb$_2$Ru$_2$O$_{6.5}$}

\begin{document}

\title[The ``ordered-ice'' oxide pyrochlore \prosix]
{Reverse Monte Carlo neutron scattering study of the ``ordered-ice'' 
oxide pyrochlore \prosix}

\author{Daniel P. Shoemaker}
\address{Materials Science Division, Argonne National Laboratory\\
Argonne, IL, 60439, USA}
\ead{dshoemaker@anl.gov}

\author{Anna Llobet}
\address{Lujan Neutron Scattering Center, Los Alamos National Laboratory\\
MS H805, Los Alamos, NM 87545, USA}
\ead{allobet@lanl.gov}

\author{Makoto Tachibana}
\address{National Institute for Materials Science\\
Namiki 1-1, Tsukuba, Ibaraki 305-0044, Japan} 
\ead{TACHIBANA.Makoto@nims.go.jp}

\author{Ram Seshadri}
\address{Materials Department and Materials Research Laboratory\\
University of California, Santa Barbara, CA, 93106, USA}
\ead{seshadri@mrl.ucsb.edu}

\date{\today}

\maketitle 

\begin{abstract}
We employ high-resolution total neutron scattering in conjunction with
reverse Monte Carlo simulations to examine,
in a detailed and unbiased manner, the crystal 
structure of the vacancy-ordered oxide pyrochlore \pro\/ in light of its 
structural analogy with proton-ordering in the structures of ice. We find that 
the vacancy and the O$^\prime$ ion are completely ordered, and that the average
structure in the $F\bar43m$ space group describes the vacancy ordering 
precisely. We complement these results with an examination of the Pb$^{2+}$ 
lone pair network using density functional electronic structure calculations, 
and a comparison of the low-temperature lattice-only heat capacity of \pro\/ 
with that of other related pyrochlores.
\end{abstract}


\section{Introduction}

In recent years, pyrochlore\cite{subramanian_oxide_1983} 
A$_2$B$_2$O$_6$O$^\prime$ compounds (also described
compactly as A$_2$B$_2$O$_7$) have been explored -- in cases where A is a 
rare-earth ion and B is a transition element -- for a number of 
reasons ranging from their unusual magnetic ``spin-ice'' ground 
states\cite{ramirez_zero-point_1999} and their anomalous Hall effect
behavior\cite{taguchi_berry_2001} to possible realization of 
magnetic monopoles,\cite{morris_dirac_2009} and because certain pyrochlore
compositions, notably Y$_2$Ir$_2$O$_7$, can display electron correlation 
in the presence of spin-orbit coupling.\cite{y2ir2o7}

When pyrochlores are prepared with lone-pair active ions such as Pb$^{2+}$ 
or Bi$^{3+}$ on their A sites, rather than rare-earth or alkaline earth ions,
one can expect interesting structural effects.
An example of this is the grotesquely complicated low-temperature structure of
Bi$_2$Sn$_2$O$_7$ and its numerous polymorphs that form upon
heating.\cite{evans_bi2sn2o7,shannon_polymorphism_1980} Hector and 
Wiggin\cite{hector_synthesis_2004} reported a low-temperature preparation of
\bto\/ in 2004; a compound which prefers to disorder extensively and locally, 
and remains cubic down to a temperature of 2\,K rather than undergo a phase 
transition to a coherently disordered ground state. This observation 
led to the suggestion that just as the pyrochlore lattice hinders coherent
magnetic ordering\cite{ramirez_strongly_1994} so too possibly are coherent
lone-pair induced displacements hindered in compounds such as \bto\/
and related 
pyrochlores\cite{avdeev_static_2002,melot_displacive_2006,henderson_structural_2007} 
a phenomenon dubbed ``charge-ice''\cite{seshadri} in analogy with problems of 
proton ordering in the structure of ice-$I_c$ and 
ice-$I_h$.\cite{bernal,pauling} The 
low-temperature thermal properties\cite{melot_large_2009} and detailed
scattering studies \cite{shoemaker_atomic_2010} have supported the  
validity of this suggestion.

In contrast to A$_2$B$_2$O$_6$O$^\prime$ pyrochlores, which crystallize
in the $Fd\bar3m$ space group and whose O$^\prime$ sites describe a 
diamond lattice, in \pro\ (average structure reported from
Rietveld refinements of neutron\cite{beyerlein_neutron_1984} and
synchrotron X-ray diffraction\cite{tachibana_electronic_2006}) there is 
ordering of O$^\prime$ and a vacancy (depicted by $\square$) so that the 
compound crystallizes in the $F\bar43m$ subgroup of $Fd\bar3m$, with
O$^\prime$ and $\square$ forming a zinc blende lattice. The complete
ordered structure in space group $F\bar43m$ can be written 
A$_2$B$_2$O$_{12}$O$^\prime$$\square$. In contrast to A$_2$B$_2$O$_6$O$^\prime$
where the diamond O$^\prime$ lattice has an A ion between every two 
O$^\prime$, in A$_2$B$_2$O$_{12}$O$^\prime$$\square$, each A ion has an
O$^\prime$ and a $\square$ as neighbors, at least in the perfectly ordered
structure (slightly further are six O neighbors in the $B_2$O$_6$ sublattice).

\begin{figure}
\centering\includegraphics[width=0.8\columnwidth]{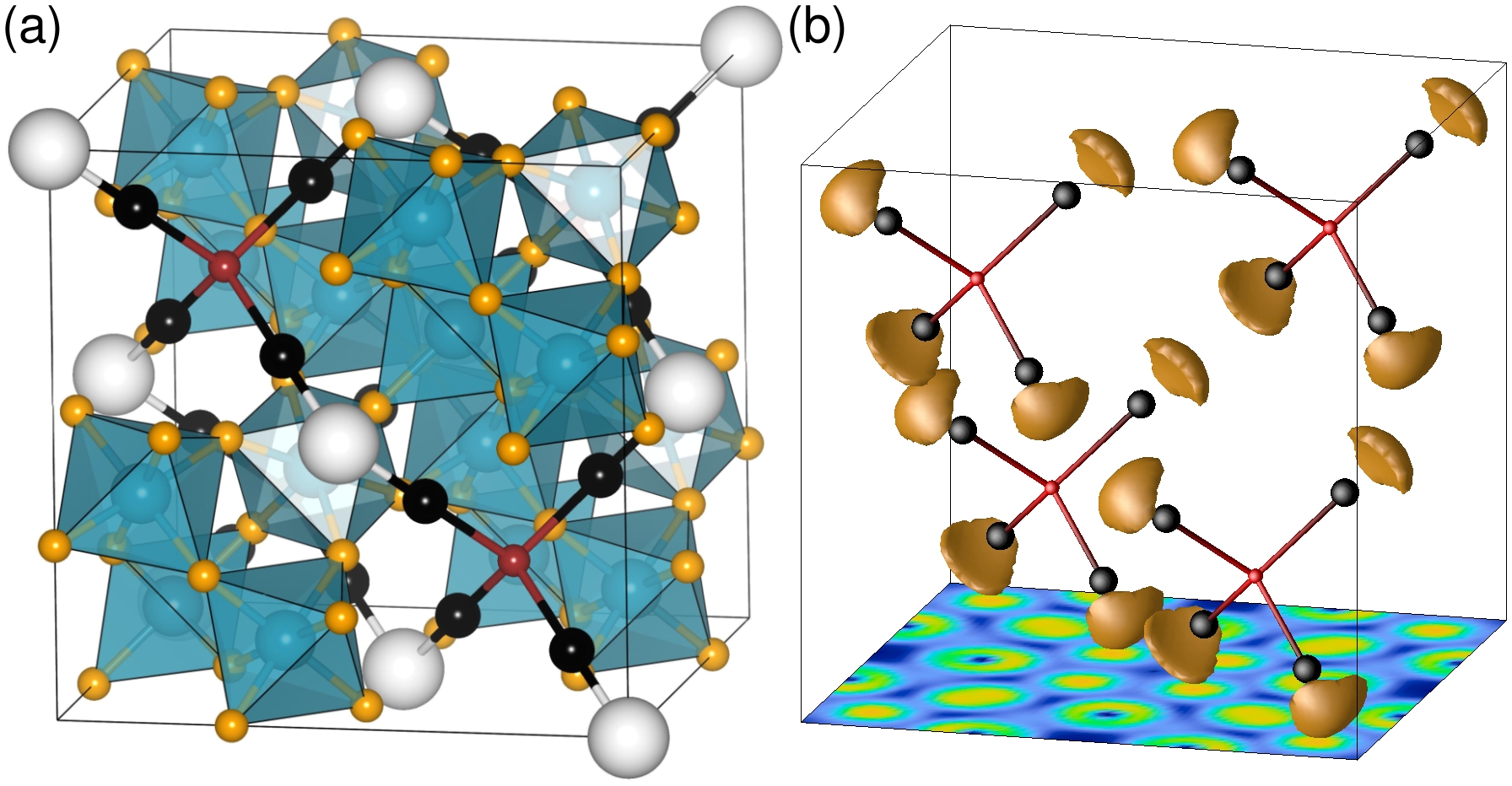} \\
\caption{(Color online) (a) The unit cell of \pro\/ consisting of 
corner-sharing networks of RuO$_6$ octahedra and isolated O$^\prime$Pb 
tetrahedra. The ordered vacancies ($\square$) are in the $4a (0,0,0)$ 
positions and are indicated using large white spheres.
(b) The lone-pairs on the Pb$^{2+}$ visualized using electron localization 
function at an isosurface of 0.80. For clarity, Ru and O are not displayed in 
(b).}
\label{fig:structure}
\end{figure}

The analogy with ice arises in the following manner: the topology of 
A$_2$O$^\prime$ in the pyrochlore structure is analogous to the topology
of H$_2$O in ice $I_c$, with the protons in the latter obeying the 
``2-in, 2-out'' Bernal-Fowler rules\cite{bernal} in a completely random way
that results in residual entropy.\cite{pauling,giauque} If the title
compound were fully ordered, the A$_2$(O$^\prime\square$)$_{0.5}$ network
would also be completely ordered with the A ions (Pb$^{2+}$) choosing to
all bond to O$^\prime$ or to $\square$ in a ``4-in 0-out'' fashion. Hence
our use of the term ``ordered-ice''. Figure\,\ref{fig:structure}(a) displays 
the crystal structure of completely vacancy-ordered \pro. The vacancy sites
are displayed as large white spheres in order to emphasize the 
Pb$^{2+}_2$(O$^\prime\square$)$_{0.5}$ network in the structure. 

In this contribution, we employ high-resolution neutron scattering tools based 
on analysis using the reverse Monte Carlo 
method\cite{mcgreavy,tucker_rmcprofile_2007} and least-squares refinements
 to establish, in a structurally
unbiased manner, the extent of ordering between O$^\prime$ and $\square$.
We find that the average structural description in the $F\bar43m$ space
group of completely ordered vacancies is consistent with the results of the
reverse Monte Carlo simulation of total scattering. An interesting
consequence of the ordered vacancies is that the stereochemically active
lone pairs on the Pb$^{2+}$ ions have a location to position themselves in 
a completely coherent fashion as we shall show. We use this 
study to emphasize the importance of employing \textit{both} the 
Bragg scattering intensity as well as the pair distribution function, 
$G(r)$ in the reverse Monte Carlo analysis. Finally, we examine the lattice 
part of the low temperature heat capacity for local Einstein modes that have
been observed in other pyrochlore compounds with incoherent lone 
pairs.\cite{melot_displacive_2006,shoemaker_arxiv} 

\section{Experimental and computational Methods}

The sample for neutron scattering was obtained by careful crushing and
grinding of \pro\ single crystals, whose average structure by synchrotron 
X-ray Rietveld refinement, and whose electrical transport properties
have been previously reported.\cite{tachibana_electronic_2006} 
Time-of-flight (TOF) total neutron scattering was carried out at the NPDF
instrument at Los Alamos National Laboratory at 300\,K and 15\,K. Rietveld
refinement was performed with the \textsc{EXPGUI} frontend for
\textsc{GSAS}.\cite{toby_expgui_2001}. Extraction of the PDF $G(r)$ with 
\textsc{PDFGetN}\cite{peterson_pdfgetn_2000} used $Q_{max}$ = 35\,\AA$^{-1}$. 
Reverse Monte Carlo simulations were run using \textsc{RMCProfile} version 6
\cite{tucker_rmcprofile_2007} and a $6\times6\times6$ unit cell with 18144
atoms. These simulations were simultaneously constrained by the Bragg profile
and the $G(r)$. Multiple runs were performed and averaged when necessary to 
obtain the most accurate representation of the fit to data.
First principles electronic structure visualization made use of the
linear muffin-tin orbital method within the atomic sphere approximation using 
version 47C of the Stuttgart TB-LMTO-ASA program.\cite{jepsen}
More details of such calculations in Pb and Bi-based pyrochlore oxides can
be found in reference \cite{seshadri}.

\section{Results and discussion}

\begin{figure} 
\centering\includegraphics[width=0.6\columnwidth]{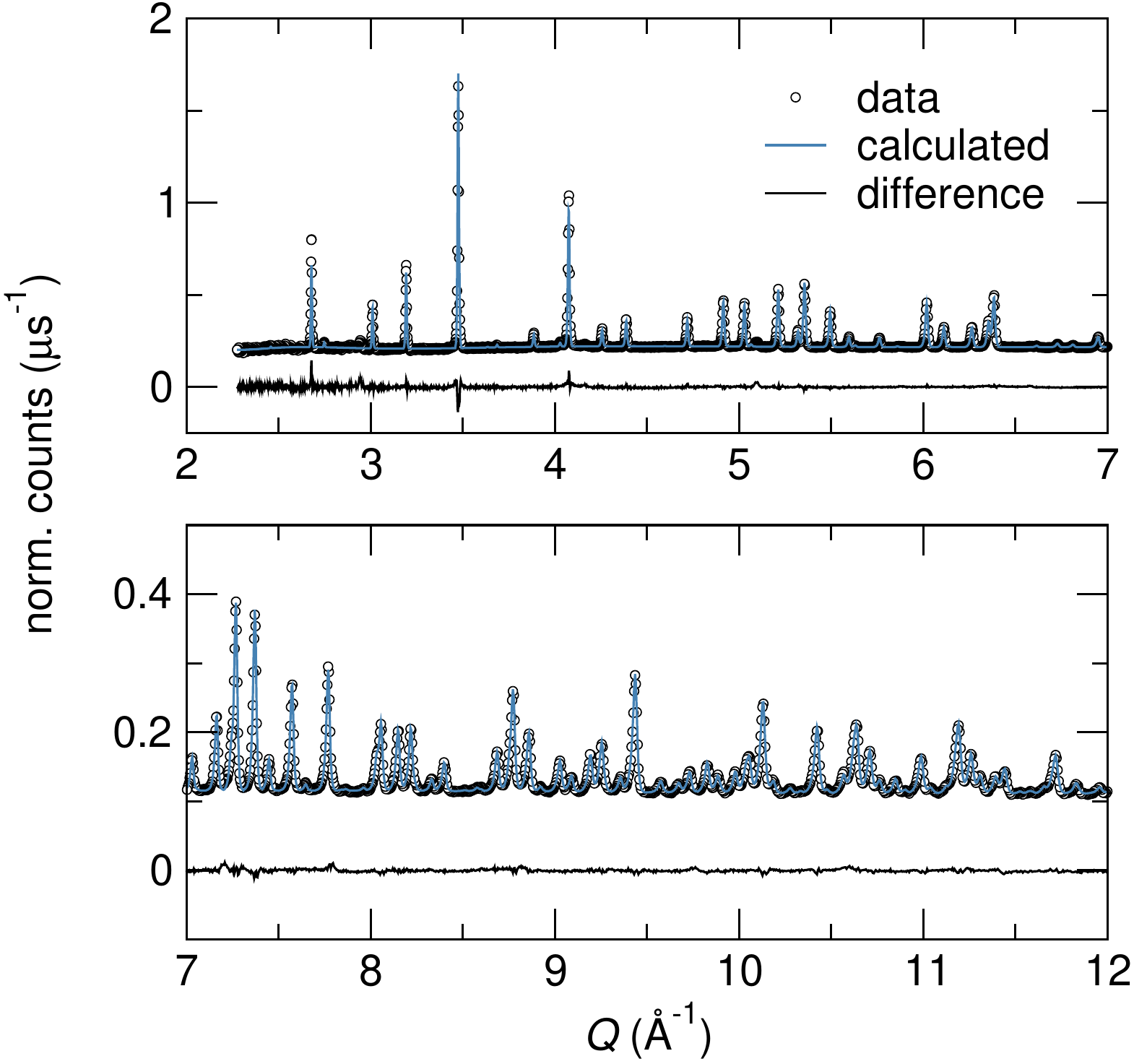} \\
\caption{(Color online) The neutron TOF Rietveld refinement for \prosix\
at $T$ = 15\,K confirms the $F\overline{4}3m$ space group. The data is 
presented in two panels to emphasize the $Q$-range and the quality of the fit.
\label{fig:rietveld}}
\end{figure}

\begin{table}
\caption{\label{tab:structure} 
Average structural parameters for \pro\ obtained from Rietveld refinement 
of TOF neutron diffraction data in the space group $F\overline{4}3m$ (No. 216).
$a$ = 10.25017(2)\,\AA/ at 300\,K and $a$ = 10.22775(2)\,\AA\/ at 15\,K. 
All sites refine to full occupancy.}
\centering
\begin{tabular}{llllll}
\hline\hline
\multicolumn{6}{c}{$T$ = 300\,K}\\
\hline \hline
atom       & site  & $x$       & $y$        & $z$       & $U_{iso}$ (\AA$^2$)\\
\hline \hline
Pb         & $16e$ & 0.8766(1) & $x$        & $x$       & 0.0117(1)\\
Ru         & $16e$ & 0.3748(1) & $x$        & $x$       & 0.0057(1)\\
O1         & $24f$ & 0.3053(2) & 0          & 0         & 0.0105(4)\\
O2         & $24g$ & 0.4502(2) & $\frac14$  & $\frac14$ & 0.0082(3)\\
O$^\prime$ & $4d$  & $\frac34$ & $\frac34$  & $\frac34$ & 0.0089(3)\\
$\square$  & $4a$  & 0         & 0          & 0         & --\\
\hline \hline
\multicolumn{6}{c}{$T$ = 15\,K}\\
\hline \hline
atom       & site  & $x$       & $y$        & $z$       & $U_{iso}$ (\AA$^2$)\\
\hline \hline
Pb         & $16e$ & 0.87725(8)& $x$        & $x$       & 0.00377(8)\\
Ru         & $16e$ & 0.3747(1) & $x$        & $x$       & 0.00299(8)\\
O1         & $24f$ & 0.3039(1) & 0          & 0         & 0.0063(2)\\
O2         & $24g$ & 0.4504(1) & $\frac14$  & $\frac14$ & 0.0042(3)\\
O$^\prime$ & $ 4d$ & $\frac34$ & $\frac34$  & $\frac34$ & 0.0051(2)\\
$\square$  & $4a$  & 0         & 0          & 0         & --\\
\hline \hline
\end{tabular}
~\\
\end{table}

The result of Rietveld refinement using neutron TOF scattering is shown in
Fig.\,\ref{fig:rietveld}. No impurities are present, and the $F\overline{4}3m$
unit cell provides an excellent fit, with details into the high-Q range
reproduced past $Q$ = 12\,\AA$^{-1}$. This model contains Pb atoms that are
shifted off the ideal pyrochlore $A$ site into a $16e$ position at ($x, x,
x$). The O$^\prime$ anions occupy $4d$ sites. Details of the structure obtained
from Rietveld refinement of the 300\,K and 15\,K data are displayed in 
Table\,\ref{tab:structure}. All atoms refine to their full occupancy and
all the isotropic thermal parameters from the refinement refine to 
acceptable values.

With the ideal, average structure, density functional calculations can be 
performed, and used to locate the lone pairs on Pb$^{2+}$\cite{seshadri_jmc} 
using the electron localization function (ELF).\cite{becke,silvi} 
Figure\,\ref{fig:structure}(b) displays the lone pair network, visualized for
an isosurface value of ELF = 0.80. The lone pairs on Pb$^{2+}$ are seen to
coherently dispose themselves in the direction of the $\square$. It is 
interesting to note that while lone pairs are known to possess a volume
that is comparable to that of oxide anion, the lone pair to nucleus 
(of Pb$^{2+}$) distance is significantly smaller than the distance from
Pb$^{2+}$ to O$^\prime$. Hence, the lone pairs are not actually located in the vacant
$\square$ site at $4a (0,0,0)$. This is in keeping with the empirical findings
of Galy, Andersson and others.\cite{galy_stereochimie_1975} 
It is instructive to compare this ordered
lone pair arrangement with what is seen in ``charge-ice'' \bto. In 
\bto, which has no vacant site, the lone pairs are obliged to displace 
perpendicular to the O$^\prime$--A--O$^\prime$ axis, and they do so
incoherently, with profound structural and thermodynamic 
consequences.\cite{seshadri,melot_large_2009,shoemaker_atomic_2010}

\begin{figure} 
\centering\includegraphics[width=0.6\columnwidth]{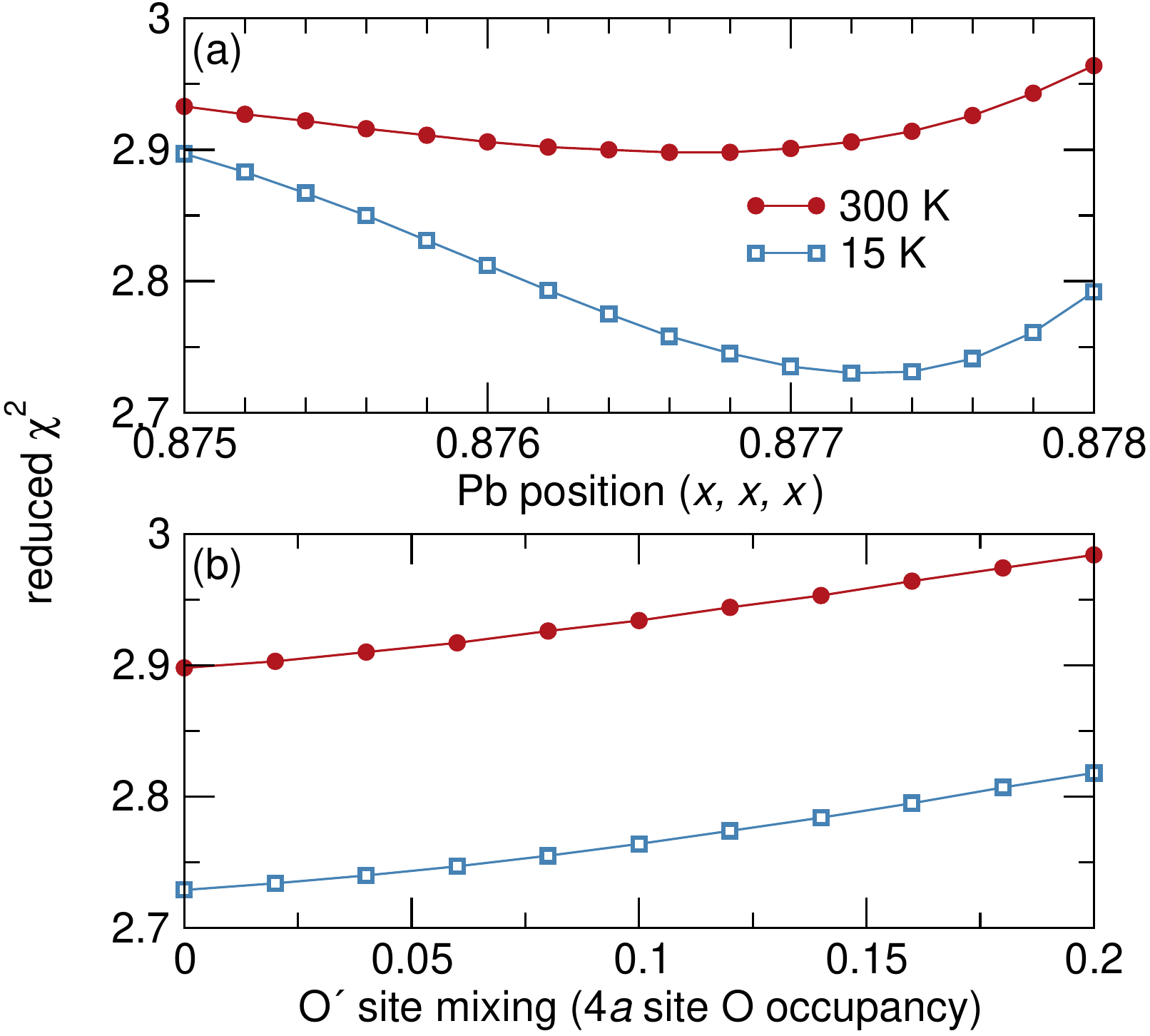} \\
\caption{(Color online) In (a), moving Pb off of its refined position at $x$ =
0.8766 results in a poorer fit, seen from the poorer fit (larger $chi^2$) 
for data acquired at 15\,K and 300\,K. 
In (b), O$^\prime$ occupancy perfectly ordered (left) and
increased to simulate disorder spilling onto the vacant sites. Again, even
small amounts of disordered O$^\prime$ give a poorer Rietveld refinement.
\label{fig:nudge}} \end{figure}

Manually altering the Pb positions or O$^\prime$ ordering in the model reveals
that both of these values are strongly constrained by the data. In
Fig.\,\ref{fig:nudge} the effect on the goodness-of-fit $\chi^2$ is plotted as
a function of Pb displacement. For an ideal pyrochlore cell, the Pb would lie
at the left-most end of the plot with $x$ = $\frac78$ = 0.875. Refinements at 
$T$ = 15\,K and 300\,K both suggest Pb positions near 0.877, 
which are about 0.04 and 0.03 \AA\ displaced, respectively,
off their ideal sites and away from the nearby O$^\prime$.

Moving fractional O$^\prime$ occupancy onto the $4a$ vacant $\square$
 site in our model is
equivalent to introducing site disorder, where having full occupancy on both
sites would mimic the O$^\prime$ sublattice  of the full pyrochlore structure.
The effect of increasing O$^\prime$ occupancy on the $\square$ site 
(with corresponding decrease at the O$^\prime$ $4d$ position) is seen
in Fig.\,\ref{fig:nudge}(b). Again, any occupancy on $4a$ results in an 
increased $\chi^2$, implying that Bragg scattering supports O$^\prime$ being
fully ordered, even at 300\,K. This strong proclivity to order the $O^\prime$
and, in turn, the Pb displacements, is seen in related solid solutions 
Pb$_{2-x}Ln_x$Ru$_2$O$_{7-y}$ ($Ln$ = Nd, Gd) which can tolerate
only a small amount ($x < 0.2$) of $A$-site substitution before the
disorder drives formation of a  separate $Fd\overline{3}m$
phase.\,\cite{kobayashi_synthesis_1995}

The quality of the fit to the Rietveld-refined structure and the stabilities
of the  Pb position and O$^\prime$ ordering provide evidence that \pro\ appears
structurally to be consistent with the unit cell, unlike the full-O$^\prime$
pyrochlores such as \bto\/ where the difficulty in accomodating the lone 
pairs results in incoherent structural disorder and off-centering. 
In \bto, local structure analysis reveals details of the short-range structure 
that are difficult to determine from average structure probes that rely on 
Bragg scattering alone. In a system such as \pro\ where structural frustration 
should be removed by O$^\prime$ ordering, it is of interest to ask whether the
local structure exhibits any hallmarks of a disordering that is not seen 
in the Bragg scattering/Rietveld analysis. 

\begin{figure}
\centering\includegraphics[width=0.6\columnwidth]{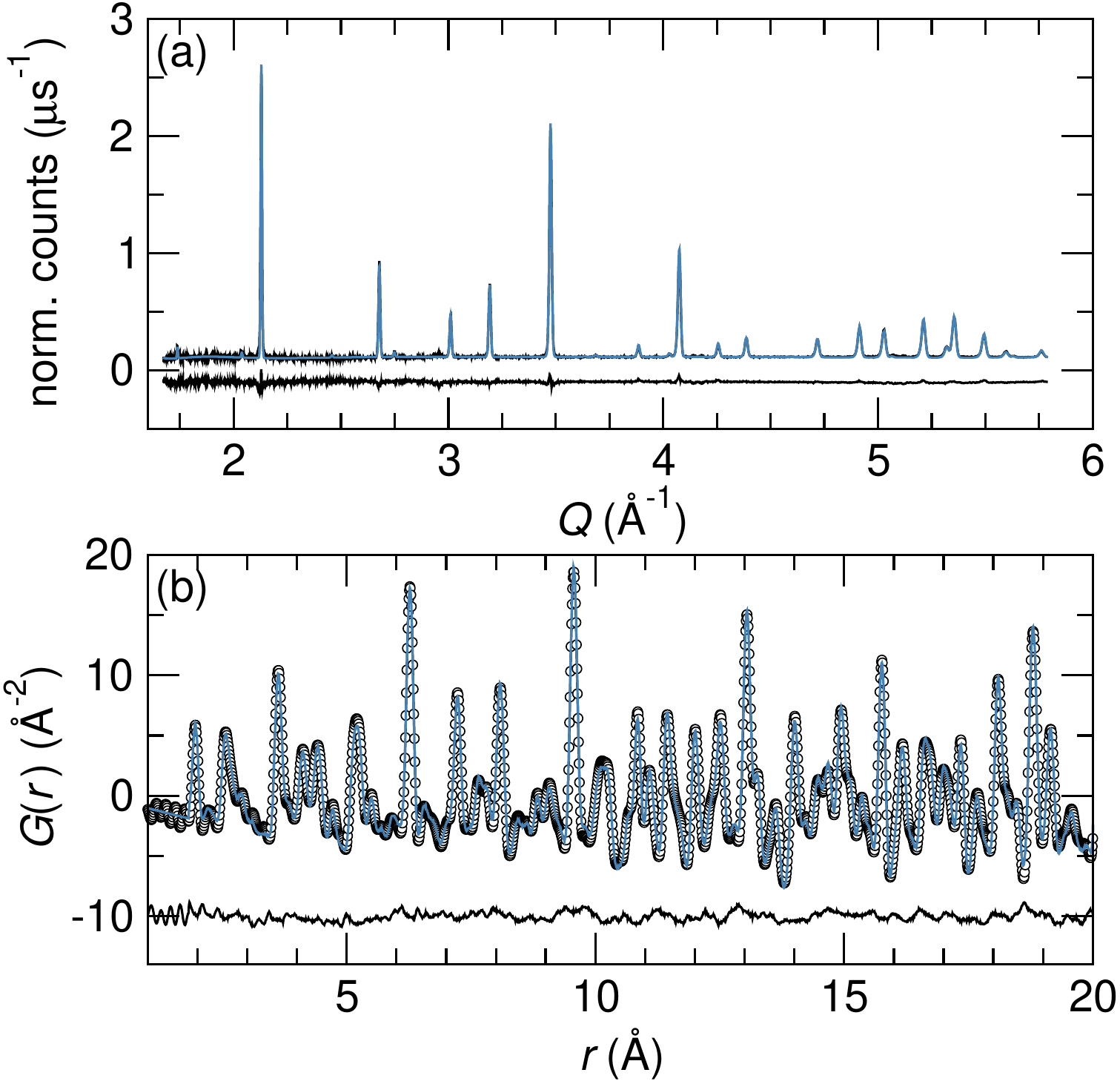}\\
\caption{ (Color online) Reverse Monte Carlo (RMC) fits to the Bragg
profile (a) and $G(r)$ (b) obtained from neutron total scattering at 15\,K.
Pair correlations are extracted from these RMC supercells.}
\label{fig:rmcfits}
\end{figure}

Reverse Monte Carlo fits to the neutron TOF total scattering are shown in
Fig.\,\ref{fig:rmcfits}. The Bragg profile retains the long-range periodicity
and average atomic displacement parameters of the structure, while the $G(r)$
constrains pairwise distances and reproduces the local structure. By
simultaneously fitting both datasets, we are left with a large supercell that
provides a snapshot of the material. Structural details of this model can be
analyzed statistically to determine the precise local tendencies supported by
the experimental data. 

\begin{figure}
\centering\includegraphics[width=0.6\columnwidth]{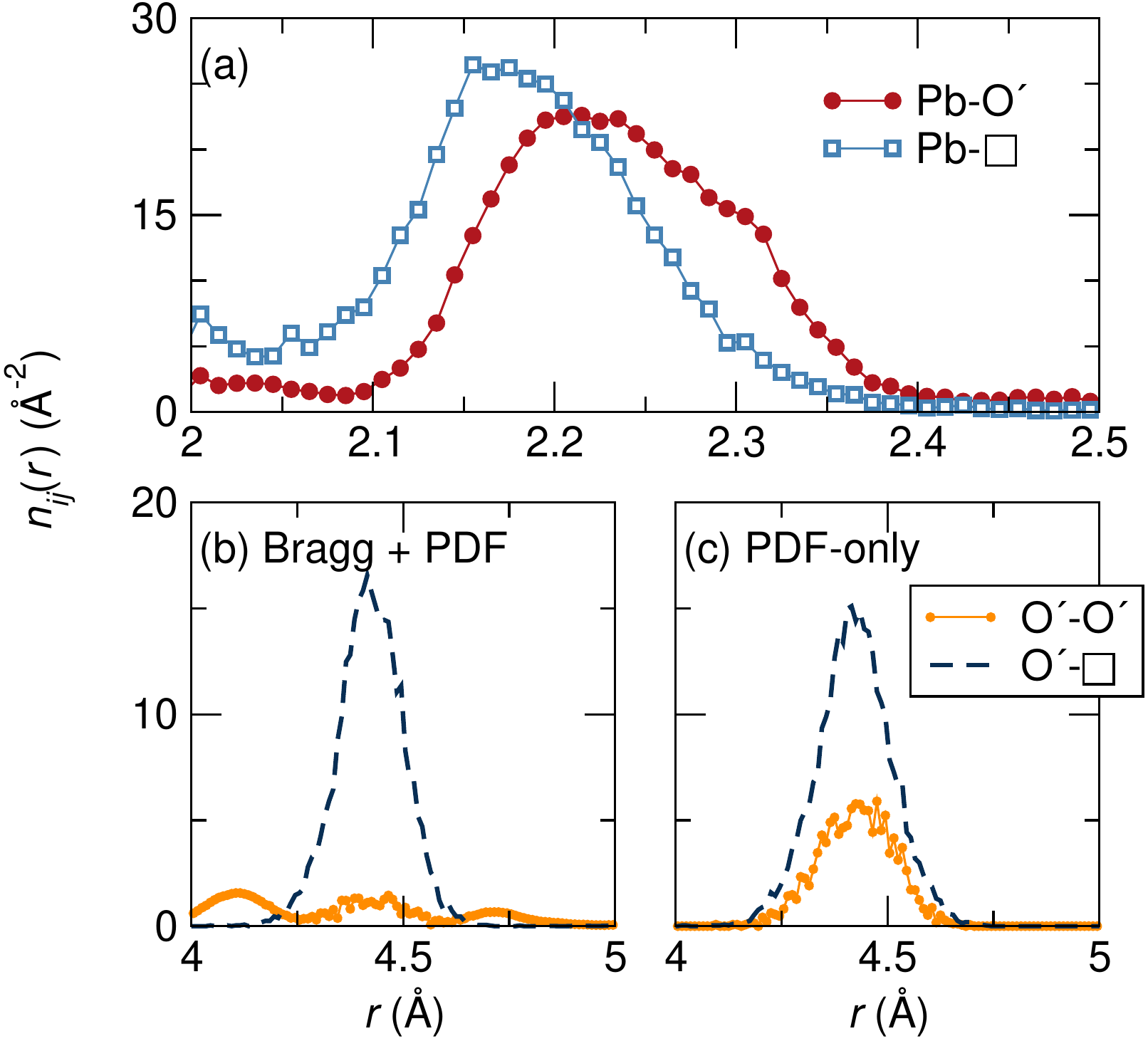}\\
\caption{ (Color online) In (a), pair correlations from the RMC supercell show
the Pb--O$^\prime$ and Pb--vacancy distances. The peaks are offset, indicating
a shift of Pb toward O$^\prime$. In (b), O$^\prime$--O$^\prime$ and
O$^\prime$--vacancy pair correlations show that O$^\prime$ are ordered when
RMC simulations are constrained by the $G(r)$ and Bragg profile. A fit
unconstrained by Bragg scattering (c) becomes partially disordered on the
O$^\prime$--vacancy sublattice--the $G(r)$ is insufficient to reproduce the
ordered distribution via RMC.}
\label{fig:partials}
\end{figure}

The partial pair distribution functions in Fig.\,\ref{fig:partials}(a) display
the  Pb--O$^\prime$ distances in the unit cell $n_{Pb-O^\prime}(r)$ and the
distances of Pb to the vacant site $n_{Pb-\square}(r)$. The shift versus $r$ shows
that Pb--O$^\prime$  distances are longer, so Pb nuclei have displaced off
their ideal site toward the vacant $\square$ position.

Partial pair distribution functions can also reveal ordering on the O$^\prime$
sublattice when vacancy ``atoms'' with zero scattering cross-section are
placed on the vacant O$^\prime$ positions. The vacancies are freely allowed to
swap with O$^\prime$ during the RMC simulation. For a simultaneous
fit to Bragg peaks and the $G(r)$, ordering is retained: the $r = 4.45$ \AA\
nearest-neighbor distance between occupied and vacant O$^\prime$ sites has no
intensity for the $n_{O^\prime-O^\prime}(r)$ distribution in
Fig.\,\ref{fig:partials}(b), indicating that no two O$^\prime$ atoms share
neighboring sites. However, if the RMC simulation is performed as a fit to the
$G(r)$ only (and the Bragg profile is ignored) then the O$^\prime$ ordering
begins to melt and O$^\prime$ can be found on the vacancy sites, as seen in
Fig\,\ref{fig:partials}(c). Both models are, at some level, ``correct''
because they reproduce the respective experimental data. However, the data
guiding the $G(r)$-only fit is incomplete. The
O$^\prime$ ordering prescribed by the Bragg pattern and verified in
Fig.\,\ref{fig:nudge}(b) is unambiguous, so the fully ordered case is confirmed.
Because RMC is a stochastic technique, it reveals that the $G(r)$ alone
does not contain sufficiently robust information about the O$^\prime$
occupancy to  prevent O$^\prime$ from spilling onto disordered sites. Only
when the additional Bragg constraint is applied is the fully ordered model
reproduced from the data.

\begin{figure}
\centering\includegraphics[width=0.6\columnwidth]{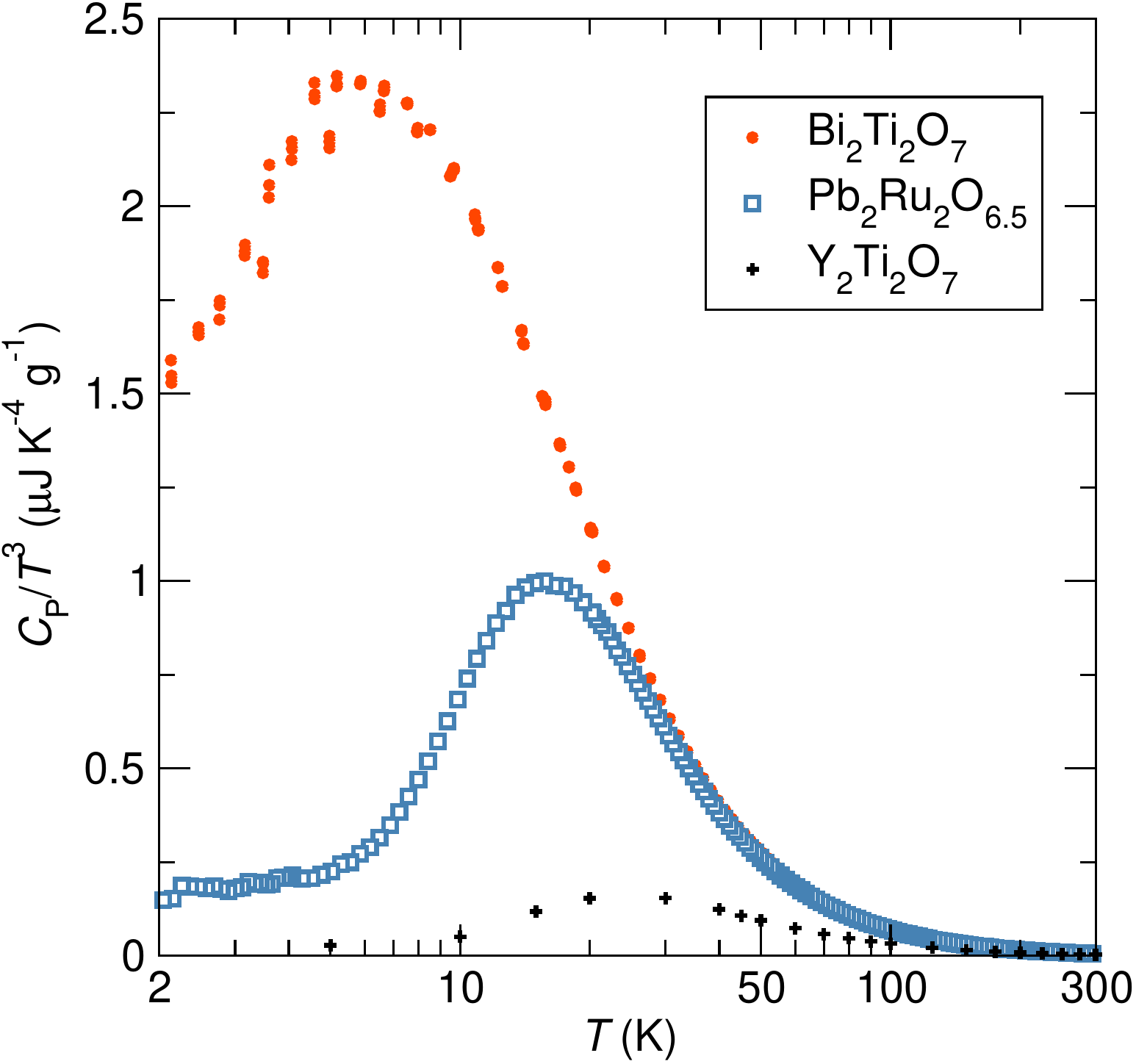} \\
\caption{(Color online) Lattice-only heat capacity, with the electronic
contribution subtracted, of \pro, displayed in a manner that emphasizes
Einstein-like features at low temperatures. For comparison, the similarly
scaled heat capacity of ``charge-ce'' \bto\/ (from reference 
\cite{melot_large_2009}) and of pyrochlore Y$_2$Ti$_2$O$_7$ 
(taken from reference \cite{white_yto}) which does not have lone pairs, is also
displayed.}
\label{fig:cp}
\end{figure}

Finally, it is of interest to see whether the ordered arrangement of Pb$^{2+}$
ions and their associated lone pairs result in distinctly different
low-temperature heat capacity behavior in \pro\/ when compared to a compound
with disordered lone pairs such as \bto. We display such a comparison
in Fig.\,\ref{fig:cp}. The scheme used to display the scaled heat capacity,
$C_p/T^3$ \textit{vs.} $T$ allows local Einstein-like modes to be 
disinguished,\cite{pohl} and the temperature of the peak(s) can be read off
as energies of the Einstein mode. Liu and L\"ohneysen\cite{liu} have suggested
a general scaling law wherein the lower the temperature at which $C_p/T^3$ 
displays a maximum, the larger the peak height as well. They have made
such correlations for a number of amorphous and crystalline systems and 
\bto\ (with its hightly disordered Bi$_2$O$^\prime$ 
network\cite{shoemaker_atomic_2010}) displays, of all studied crystalline
systems, one of the lowest peak temperatures and largest peak amplitudes.
In contrast, we see from Fig.\,\ref{fig:cp} that in \pro, (using published
data from reference \cite{tachibana_electronic_2006} after subtracting the
electronic $\gamma T$ contribution) the peak occurs at significantly higher 
temperatures and is significantly smaller in amplitude. It is nonetheless
larger than what is seen for Y$_2$Ti$_2$O$_7$\cite{white_yto} which 
has no lone pairs. The question is why, in a completely ordered structure,
is there an Einstein-like mode at all? The suggestion made recently by
Safarik \textit{et al.}\cite{hundley} is that Einstein-like modes
appear even in crystalline materials when van Hove peaks in the phonon
densities of state cross the chemical potential. In any event, the trend 
seen in Fig.\,\ref{fig:cp} is consistent with the finding that \pro\/ has a 
well-ordered network of A-site atoms unlike \bto.

\section{Conclusions}

Our study of the ordered-ice pyrochlore \pro\ has found that Rietveld
refinements dictate O$^\prime$-$\square$ occupancy to be fully ordered,
and Pb$^{2+}$ is displaced toward O$^\prime$ (0.04 \AA\ at $T$ = 15 K).
This ordering implies that \pro\ does not possess the same geometric
frustration seen in fully-occupied  $A_2B_2$O$_7$
pyrochlores with lone-pair active cations on the $A$ sites. To compare
\pro\ to those systems, where local structure analysis provide a view of
incoherent distortions, we performed RMC simulations which revealed the
presence of \emph{coherent} Pb${^2+}$ off-centering only. Simulataneous
fits to the Bragg intensity and $G(r)$ are required to reproduce the vacancy
ordering, which is not strongly constrained by the $G(r)$ alone.
First-principles calculations and heat capacity measurements corroborate
to show that the O$^\prime$-$\square$ ordering provides an outlet for (and perhaps
is constrained by) the pointing of lone pairs into vacant $4a$ positions.

\section{Acknowledgments}

We thank Joan Siewenie for assistance with data collection at NPDF.
DPS and RS gratefully acknowledge support from the UCSB-LANL Institute for 
Multiscale Materials Studies, and from the National Science Foundation 
(DMR 0449354). DPS additionally acknowledges
work at Argonne National Laboratory supported by the U.S. DOE, Office of
Science, under Contract DE-AC02-06CH11357. This work made use of MRL Central 
Facilities, supported by the MRSEC Program of the NSF 
(DMR05-20415), a member of the NSF-funded Materials Research Facilities
Network (www.mrfn.org). NPDF at the Lujan Center at Los Alamos Neutron 
Science Center is funded by the DOE Office of Basic Energy Sciences and 
operated by Los Alamos National Security LLC under DOE Contract 
DE-AC52-06NA25396. RMC simulations were performed on the Hewlett Packard 
QSR cluster at the CNSI-MRL High Performance Computing Facility. 

~\\

\bibliography{pro}

\end{document}